\documentclass{aa}
\usepackage{epsf,url}
\usepackage{amssymb,amsfonts,amstext,amsgen,amsopn,amsxtra,indentfirst,lscape,times,rotating}
\usepackage{longtable}
\usepackage{pdflscape}
\usepackage{graphics}
\usepackage{keyval,ulem}
\usepackage{trig}
\usepackage[dvips]{color}%\usepackage{colourslide}
\usepackage{dcolumn}% Align table columns on decimal point
\usepackage{bm}% bold math
\newcommand\ignore[1]{} 
\def\beq{\begin{equation}}
\def\eeq{\end{equation}}
\def\bey{\begin{eqnarray}}
\def\eey{\end{eqnarray}}

\def\mnras{MNRAS}
\def\apj{ApJ}

\def\apjs{ApJ}
\def\apjl{ApJ}

\def\araa{ARAA}

\def\aap{A \& A}
\def\aj{AJ}

\def\aap{Astron. Astrophys.}

\begin{document}

\title{Dynamical measurement of the stellar surface density of face-on galaxies}
\titlerunning{Stellar surface density of face-on galaxies}

\author{G. W. Angus\fnmsep\inst{1}\thanks{E-mail: garry.angus@vub.ac.be}
\and G. Gentile\fnmsep\inst{2,1}
\and B. Famaey\fnmsep\inst{3}
}
\institute{
Department of Physics and Astrophysics, Vrije Universiteit Brussel, Pleinlaan 2, 1050 Brussels, Belgium   
\and
Sterrenkundig Observatorium, Universiteit Gent, Krijgslaan 281, 9000, Gent, Belgium
\and
Observatoire astronomique de Strasbourg, Universit\'e de Strasbourg, CNRS UMR 7550, rue de l'Universit\'e 11, 67000 Strasbourg, France
}

\date{Received date; accepted date} 

\abstract{The DiskMass survey recently provided measurements of the vertical velocity dispersions of disk stars in a sample of nearly face-on galaxies. By setting the disk scale-heights to be equal to those of edge-on galaxies with similar scale-lengths, it was found that these disks must be sub-maximal, with surprisingly low K-band mass-to-light ratios of the order of $M_\star/L_K \simeq 0.3 M_\odot/L_\odot$. This study made use of a simple relation between the disk surface density and the measured velocity dispersion and scale height of the disk, neglecting the shape of the rotation curve and the dark matter contribution to the vertical force, which can be especially important in the case of sub-maximal disks. Here, we point out that these simplifying assumptions led to an overestimation of the stellar mass-to-light ratios. Relaxing these assumptions, we compute even lower values than previously reported for the mass-to-light ratios, with a median $M_\star/L_K \simeq 0.18 M_\odot/L_\odot$, where 14 galaxies have $M_\star/L_K < 0.11$. Invoking prolate dark matter halos made only a small difference to the derived $M_\star/L_K$, although extreme prolate halos ($q>1.5$ for the axis ratios of the potential) might help. The cross-terms in the Jeans equation are also generally negligible. These deduced K-band stellar mass-to-light ratios are even more difficult to reconcile with stellar population synthesis models than the previously reported ones.}
\maketitle


\section{Introduction}
\protect\label{sec:intr}
The kinematics of disk stars perpendicular to the plane has long been acknowledged to have great importance as a probe of the surface mass density. Pioneered almost a century ago by the works of \cite{kapteyn22} and \cite{oort32} in the Milky Way, this led to a plethora of studies in the 1980s and beyond \citep{vanderkruit81a,vanderkruit81b,vanderkruit82,bahcall84,bienayme87,vanderkruit88,kuijken89,bottema93,creze98,holmberg00,siebert03,herrmann08,herrmann09a,herrmann09b,herrmann09c,vanderkruit11,bovy13,bienayme14}, culminating, for external galaxies, with the recent work from the DiskMass Survey \citep{bershady10a,bershady11,martinsson13a,martinsson13b}.

When computing the disk surface density from a measurement of the vertical gravitational field profile, the recovered surface density can be well approximated - at least close to the mid-plane - by the cylindrical Poisson equation:

\beq
\protect\label{eqn:radder}
2\pi G \Sigma(z)\equiv 2\pi G \int_{-|z|}^{|z|}\rho(z)dz=|K_z|-|z|{1 \over R}{\partial \over \partial R} (RK_R).
\eeq
Here the surface density and gravitational field profiles are general and are comprised of stars, gas, and dark matter (DM) components.
By neglecting the second term (radial derivative) of the right-hand side of Eq.~\ref{eqn:radder} and assuming no DM or gas, \cite{vanderkruit88} noted that the stellar vertical gravitational field profile of a vertical exponential stellar disk is
\beq
|K_z^{\star}|(R,z)=2\pi G\Sigma_{\star}(R)[1-\zeta_{\star}(z)]
,\eeq
where $\zeta_{\star}(z)=\exp(-|z|/h_z)$, and $h_z$ is the scale
height of the stars. Using the Jeans equation and neglecting the small cross-term $\sigma_{Rz}$, this gravitational field can be used to find the stellar vertical velocity dispersion integrated over $z$, 
\beq
\sigma_{z\rm,\star}(R)^2 = \sum\limits_{i} \sigma_{z\rm,\star}^{i}(R)^2 , 
\eeq
with
\beq
\protect\label{eqn:sig*}
\sigma_{z\rm,\star}^{i}(R)^2=\frac{1}{2 h_z}\int_{-\infty}^{\infty} \left[\int_z^{\infty}\zeta_{\star}(z')K_z^{i}(R,z')dz' \right] dz.
\eeq
Here $i$ is used to denote the component that is contributing to the stellar velocity dispersion. If $i=\star,$ then this gives the contribution to the stellar vertical velocity dispersion from the gravitational field of stars, but if $i=$~DM, then it gives the contribution to the stellar vertical velocity dispersion from the gravitational field of DM. The subscript $\star$, on the other hand, refers to the fact that we use the stellar vertical velocity dispersions as tracers of the gravitational potential, and thus only the stellar vertical distribution, $\zeta_{\star}$, is relevant.

For the pure double exponential stellar disk of \cite{vanderkruit88}, this leads simply to
\beq
\protect\label{eqn:vdk*}
\Sigma_{\star}(R)={\sigma_{z\rm,\star}^{\star}(R)^2 \over 1.5 \pi G h_z}.
\eeq
Of course, the purely stellar contribution, $\sigma_{z\rm,\star}^{\star}$ to the total vertical velocity dispersions, $\sigma_{z\rm,\star}$, is not directly measurable, therefore, assuming that the DM component is negligible near the disk-plane, but taking into account the contribution of the atomic and molecular gas (which lie in very thin disks), $\Sigma_{\rm atom}$ and $\Sigma_{\rm mol}$, the stellar surface density is most often generalized to
\beq
\protect\label{eqn:vdk}
\Sigma_{\star}(R)={\sigma_{z\rm,\star}(R)^2 \over 1.5 \pi G h_z}-\Sigma_{\rm mol}(R)-\Sigma_{\rm atom}(R).
\eeq
Dividing the derived stellar surface density by the stellar surface brightness gives the stellar mass-to-light ratio $M_{\star}/L={\Sigma_{\star}(R) \over I_{\star}(R)}$. However, we note that in reality, this $\Sigma_{\star}$ still includes a (generally assumed to be small) DM component.

This relationship of Eq.~\ref{eqn:vdk} has been used in several studies of the disk surface density of nearly face-on galaxies, such as \cite{bottema93}, \cite{gentile15} and the DiskMass Survey (\citealt{bershady10a,bershady11} and \citealt{martinsson13a,martinsson13b}; hereafter M13a and M13b) and is generally assumed to be a valid approximation. 
Noteworthy among the aforementioned studies is the DiskMass Survey, which is a multi-wavelength kinematic survey of nearly face-on galaxies. Their goal was to unambiguously dynamically measure the mass of the stellar disk by simultaneously measuring the galaxy rotation curve and the vertical velocity dispersions of the old disk stars, as well as the stellar surface brightness profile and the disk inclination. Combining the data sets allowed them to break the disk halo degeneracy (\citealt{vanalbada85,gentile04,bershady11}). The observed stellar surface brightness for each galaxy was fit with an exponential disk with scale length $h_R$. The scale height, $h_z$, was inferred from scaling relations of edge-on galaxies (\citealt{kregel02,bershady10b}). 

%Eq~\ref{eqn:vdk} gives the total surface density.
% From this, the atomic and molecular gas surface densities were subtracted to leave the stellar plus dark matter surface density. The dark matter was assumed to be negligible within the disk and thus was not considered further. 

Using the simple method outlined above, they concluded from a sample of 30 galaxies that galaxy disks are much lighter than previously thought, since the derived mass-to-light ratios were roughly half those predicted from models of stellar population synthesis (e.g. \citealt{mcgaugh14}), and that the disks were highly sub-maximal. However, with ever-decreasing uncertainties on all component parts, it is worth establishing if the approximation used above allows the stellar surface density, and thus the mass-to-light ratios, to be accurately recovered.

The central point is thus the validity of Eq~\ref{eqn:vdk}. To derive it, the second term of Eq.~\ref{eqn:radder} was neglected, which may be significant when the rotation curve is sharply varying. What is more, it is important to remember that the first term on the right-hand side of Eq.~\ref{eqn:vdk} is in principle comprised of all matter types including DM, which is not subtracted by the two following terms. While this is a reasonable assumption in the case of maximum disks, the fact that the DiskMass result implies sub-maximality renders this assumption questionable.

Since the data from the DiskMass survey are not yet publicly available, here only quantities published in M13a and M13b are used. For each of the 30 galaxies in their sample, the fitted DM halo parameters, exponential stellar disk parameters, and the derived mass-to-light ratios are used. This should provide a large enough sample to clarify the importance of the two aforementioned issues. In Sect. \ref{sec:disk} we briefly review the importance of the radial term in the Poisson equation. In Sect. \ref{sec:vvd} the stellar mass-to-light ratios are re-derived using the Jeans equation and taking into account the DM distribution's influence on the stellar velocity dispersions. In Sect. \ref{sec:conc} the conclusions are presented.

%The total surface density was established from the observed stellar vertical velocity dispersions with Eq~\ref{eqn:vdk}.  thus the stellar mass-to-light ratio was found using 

%\beq
%\protect\label{eqn:mlsimp1}
%(M_{\star}/L)={\Sigma_{\star}(R) \over I_{\star}(R)}.
%\eeq
% If the dark matter could be subtracted then the stellar mass-to-light ratio could be found as $(M_{\star}/L)={\Sigma_{\star}(R) \over I_{\star}(R)}$. 

%\cite{martinsson13a} fitted exponential disks to the galaxy photometry as well as dark matter halos to the rotation curves, both NFW (\citealt{nfw97}) and pseudo isothermal spheres. Here the focus is placed on the pseudo isothermal sphere (pISO), whose density is given by

%These two aforementioned quantities are assumed to be fixed by observations for each galaxy. The projected stellar luminosity density can be compared, for example, to the stellar surface density to give a stellar mass-to-light ratio (refs).
%The projected luminosity density can also be compared to the stellar plus dark matter surface density to give a stellar plus dark matter mass-to-light ratio
%\protect\label{eqn:mlsimp2}
%(M_{DM}/L)&=&{\Sigma_{DM}(R) \over I_{\star}(R)},
%eey

\section{Deriving the disk surface density}
\protect\label{sec:disk}

The rotation curve often significantly varies at the radii where the DiskMass Survey galaxies have measured velocity dispersions, mostly $0.5h_R<R<2h_R$, meaning that neglecting the radial term in the Poisson equation (as was done to obtain Eq.~\ref{eqn:vdk}) could lead to systematics. To recall the importance of these errors for the DiskMass galaxies, we calculate the derived dynamical surface densities for two approximations.

The first approximation to the surface density is equivalent to the approximation used by the DiskMass Survey, in the sense that both approximations neglect the radial derivatives in Eq~\ref{eqn:radder}. This is found from 

\beq
\protect\label{eqn:radder1}
2\pi G \left[\Sigma_{\star}(z)+\Sigma_{DM}(z)\right]\sim  |K_z^{\star}|+|K_z^{DM}|,\eeq
which can be evaluated at any chosen radius, $R$, and $z=3h_z$. The height of $3h_z$ ensures that 95\% of the stellar mass is accounted for and is similar, in most cases, to the 1.1~kpc that is standard for estimating the surface density of the Milky Way (see e.g. \citealt{bovy13}).

Here we include DM to make the scenario more realistic and to avoid the misunderstanding that any errors come from the Keplerian decline of the stellar contribution to the rotation curve. This requires the vertical gravitational field profile for the DM halo and a double exponential disk (see \citealt{kuijken89}). The pseudo-isothermal sphere DM halo parameters and the stellar disk scale lengths and scale heights can be found in M13a (their Table 1). The stellar mass-to-light ratio adopted here for each galaxy is simply the one derived by M13a ( their Table 6, second column). This approximation is roughly equivalent to that which gives Eq.~\ref{eqn:vdk} because that equation uses measured velocity dispersions (generated by the true double exponential disk vertical gravitational field) and an incomplete formula to derive the surface density, whereas Eq.~\ref{eqn:radder1} uses the true double exponential disk vertical gravitational field and a formula with the same flaw to derive the surface density.

\begin{figure}
\includegraphics[angle=0,width=8.50cm]{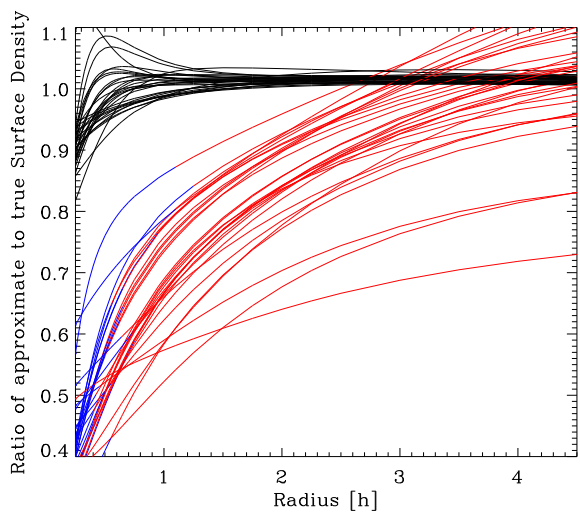}
\caption{Ratio of the approximate total surface density to the true surface density as a function of radius. Each line represents the ratio of an approximate total surface density to the true total surface density within $3h_z$ of the mid-plane for each galaxy as a function of radius. The black lines account for the radial derivatives in the Poisson equation (Eq.~\ref{eqn:radder2}) and the red lines neglect them (Eq.~\ref{eqn:radder1}). At radii where the bulge is significant, the red lines turn blue.}
\label{fig:surfd}
\end{figure}

The second approximation to the surface density uses
\bey
\protect\label{eqn:radder2}
\nonumber &&2\pi G \left[\Sigma_{\star}(z)+\Sigma_{DM}(z)\right] \cong \\
&&|K_z^{\star}|+|K_z^{DM}|-|z|{1 \over R}{\partial \over \partial R} \left[R(K_R^{\star}+K_R^{DM})\right],
\eey
which is just an expanded version of Eq.~\ref{eqn:radder} and thus accounts for the otherwise neglected radial derivatives term in the Poisson equation. The ratios of these two approximations with respect to the true surface density, $\Sigma_{\star}(z=3h_z)+\Sigma_{DM}(z=3h_z$), are plotted in Fig~\ref{fig:surfd}.

%It is then assumed that a second measurement, using stellar velocity dispersions, faithfully provides the amplitude of the total vertical gravitational field profile (including stars and dark matter), where the shape of the stellar vertical gravitational field profile is taken to be that of a double exponential disk with 

In all cases, Eq.~\ref{eqn:radder2} (black lines) recovers the actual total surface density to better than a few percent at $R>h_R/2$.  Neglecting the radial derivatives (red lines) produces large errors. Including only data beyond the bulge radius, these errors are on average 30\% at $R=h_R$ and 35\% at $R=h_R/2$. This means that the approximation of Eq.~\ref{eqn:vdk} used by the DiskMass survey under-estimates the surface density (of stars+DM, the DM being assumed negligible in the DiskMass analysis) by a significant amount. Very little difference is found if the surface density within a fixed height like 1.1~kpc is considered, or if we only treat the stellar component.

%Since for the majority of the galaxies the mass-to-light ratio found by M13a is strongly biased towards the values estimated at $h_R/2<R<h_R$, the mass-to-light ratio adopted here for the remainder of the analysis is the original, from M13a, divided by the average value of the red line from Fig~\ref{fig:surfd} in that radial range. On average, this amounts to a 30\% increase.

%Note that the discussion above relates only to the underestimation of the stellar surface density and thus stellar mass-to-light ratio, however the mass-to-light ratios derived by M13a are dark matter plus stellar mass-to-light ratios. It turns out that the stellar plus dark matter mass-to-light ratio is underestimated by a similar amount.

\section{Subtracting the dark matter contribution in sub-maximal disks}
\protect\label{sec:vvd}

The large discrepancy between the first and second approximations to the surface density provides the motivation to re-derive the stellar mass-to-light ratio with a different method that is not affected by the neglected terms in the Poisson equation and accounts for the DM close to the disk. 

%It is expected from cosmological simulations that not only should galaxies be shrouded in dark matter, but the dark matter should adiabatically contract giving it a higher density in the disk than a simple smooth component (e.g. Multidark). However, simulations that include feedback effects can reverse this contraction. Regardless, here only the smooth component is considered. In the analysis of M13a, the dark matter surface density was assumed to be negligible, however, if its contribution to the stellar vertical velocity dispersion is large, it will imply the underlying purely stellar mass-to-light ratio is lower than was found.

The influence of the DM component on the stellar mass-to-light ratios of the DiskMass Survey galaxies was previously investigated by \cite{swaters14}. There, the correction for dark matter followed the strategy of \cite{bottema93}, which depends on the estimated DM density in the mid-plane, which is not ideal. It also relies on an incomplete estimation of the stars+DM surface density from Eq~\ref{eqn:vdk}. They found that accounting for DM reduced their original average mass-to-light ratio from 0.3 to a stellar mass-to-light ratio of 0.24, but expect a larger decrease if a more rigorous approach was applied.

The approach used here is to compute the expected contribution to the stellar velocity dispersion from the DM halo gravity and the disk gravity, respectively. This is done using the Jeans equation, Eq~\ref{eqn:sig*}. We note that the contribution of the gravitational field generated by each matter type to the stellar velocity dispersion is modulated by the vertical range of the old disk stars, $\zeta_{\star}$, in Eq~\ref{eqn:sig*}. 

\begin{figure}
\includegraphics[angle=0,width=8.50cm]{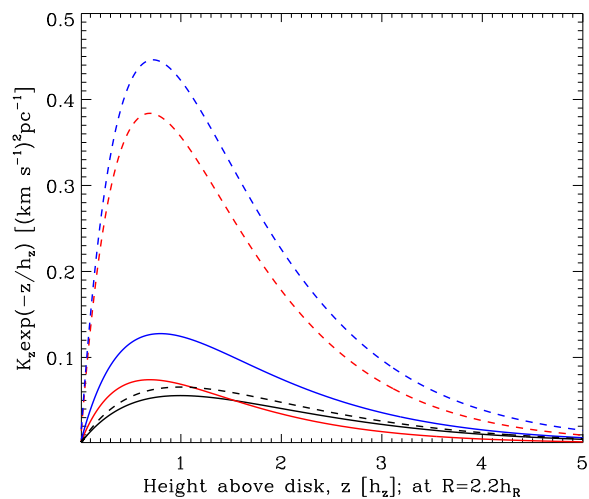}
\caption{Vertical gravitational field modulated by the vertical stellar density profile as a function of the height above the disk at $R=2.2~h_R$. The solid lines correspond to UGC~3091 and the dashed lines to UGC~3140. The red lines reflect the stellar gravitational field, the black lines reflect the dark matter, and the blue lines are the sum of the black and red lines.}
\label{fig:linlin}
\end{figure}

Although the stars typically outweigh the DM in the mid-plane, the vertical component of the DM gravitational field can often reach parity at about one scale height. Obviously, this will be even more severe when the stellar mass-to-light ratio is low. This is illustrated in Fig~\ref{fig:linlin}, which shows the weighted vertical force $\zeta_{\star}(z)K_z(z)$ for the stellar disk and DM halo components. For the stellar contribution, the  $M_{\star}/L$ from M13a is used (and hence is most often sub-maximal) and the DM halo is the pseudo-isothermal sphere fitted by M13a to the rotation curve (see their Table 6, Cols. 4 and 5). This is given for one of the more DM-dominated galaxies, UGC~3091, and one of the most disk-dominated galaxies, UGC~3140.

\begin{figure}
\includegraphics[angle=0,width=8.50cm]{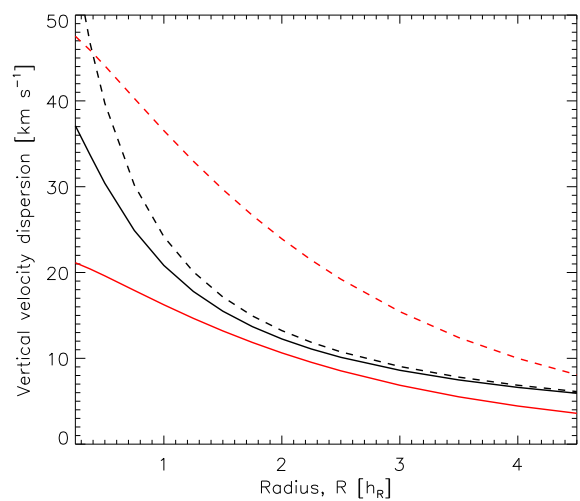}
\caption{Contribution from different components to the model vertical velocity dispersions for two galaxies UGC 3091 (solid lines) and UGC~3140 (dashed lines). The black lines correspond to the contribution to the galaxy's stellar vertical velocity dispersion from dark matter, whereas the red lines are the stellar contribution. }
\label{fig:vvd}
\end{figure}

In Fig.~\ref{fig:vvd} the contribution to the stellar vertical velocity dispersion profile from the DM gravitational field is compared with the stellar contribution. They are computed using the dark matter and stellar vertical gravitational field profile as the input to Eq.~\ref{eqn:sig*}, respectively.  For UGC~3091, the DM gravitational field dominates the induced stellar vertical velocity dispersion at all radii, and thus should certainly not be ignored. We recall that the DiskMass strategy was to measure the stellar mass-to-light ratio by assuming a {\it negligible} contribution of the DM gravitational field to the velocity dispersion. On the other hand, for UGC~3140 the DM gravitational field contribution to the stellar velocity dispersion is irrelevant at most radii.

%Note that if the quoted $M_{\star +DM}/L$ from M13b was not increased by a value of 0.1, then the total contribution to the vertical velocity dispersion would not correspond to the observed profiles and thus the importance of dark matter would be overemphasised.

\subsection{Re-deriving the stellar mass-to-light ratios}
\protect\label{sec:m2l}

The total measurable stellar vertical velocity dispersion should thus be $\sigma_{z\rm, \star}^{tot}(R)^2=\sigma_{z\rm, \star}^{\star}(R)^2+\sigma_{z\rm, \star}^{DM}(R)^2+\sigma_{z\rm, \star}^{gas}(R)^2$. In the absence of the measured stellar velocity dispersions for each galaxy, a substitute must be used. The stellar velocity dispersion data, as a function of radius, are generally well fit by an exponential profile of the form

\beq
\protect\label{eqn:m13b}
\sigma_{z\rm, \star}^{tot}(R)=\sigma_{z\rm, \star}^{tot}(0)\exp(-R/h_{\sigma}),
\eeq
and for this reason, M13b modelled each of the 30 galaxies to find the best fit $\sigma_{z\rm, \star}^{tot}(0)$ and $h_{\sigma}$ (see Table 6 of M13b). Therefore, instead of the individual stellar velocity dispersion data points, the best-fit exponential profile using Eq.~\ref{eqn:m13b} is evaluated over a particular radial interval. These `data points'  are fitted with the combined DM and stellar model velocity dispersions where the only free parameter is the stellar mass-to-light ratio. The potential of the DM halo and of the exponential disk are exact solutions of the Poisson equation, and hence take into account the radial derivative term by construction. The velocity dispersion exponential profiles from Eq.~\ref{eqn:m13b} include a typically small component of atomic and molecular gas that is not explicitly modelled here. To account for it, an amount equal to its contribution to the dynamical mass-to-light ratio in M13a (figures in their atlas A3) is subtracted from our fitted mass-to-light ratios. This is generally around 0.1.

\begin{figure}
\includegraphics[angle=0,width=8.50cm]{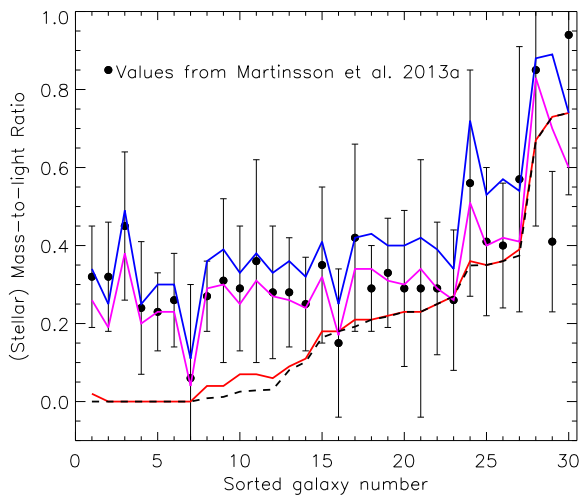}
\caption{Mass-to-light ratios derived using the four methods described in Sect. \ref{sec:m2l} for all 30 galaxies from Martinsson (2013a). Black circles, with associated error bars, are the stellar mass-to-light ratios derived by Martinsson (2013a) for all 30 galaxies of the DiskMass Survey. The violet lines (case 1 from Sect. \ref{sec:m2l}) are the mass-to-light ratios we found using the DiskMass Survey method (Eq.~\ref{eqn:vdk}). The blue lines (case 2 from Sect. \ref{sec:m2l}) are the mass-to-light ratios found while accounting for the second term in the Poisson equation (Eq.~\ref{eqn:radder}), but not the DM. The red lines (case 3 from Sect. \ref{sec:m2l}) are the fitted stellar mass-to-light ratios after accounting for the second term in the Poisson equation and the dark matter needed to fit the original rotation curve (using the original mass-to-light ratio). The dashed black line (case 4 from Sect. \ref{sec:m2l}) goes one step further and identifies the increased dark matter density required to offset the decreased stellar contribution to the rotation curve from the red line. The actual stellar mass-to-light ratio is then found iteratively. Each line has had the mass attributed to the molecular and atomic gas subtracted. Negative stellar mass-to-light ratios are fixed to zero.}
\label{fig:ml}
\end{figure}

The default radial interval for comparison is [0.5-3]$\times h_R$, where the majority of the DMS galaxies have measured stellar velocity dispersions. Each radius in the interval is given the same weighting. In practice this means that we compute the $\chi^2$ between the model and observed stellar vertical velocity dispersions at several discrete radii in the interval. This is slightly different to M13a since their error bars scale with radius. 
We compute the mass-to-light ratio for four scenarios and plot them in Fig~\ref{fig:ml}:
\begin{enumerate}[(i)]

\item The first case is analogous to the DiskMass approach (violet lines), where both the contribution of the DM gravitational field and the second term of Eq.~\ref{eqn:radder} are neglected. In this case, the model vertical velocity dispersion is found from Eq.~\ref{eqn:vdk}, which allows us to deduce the best-fit mass-to-light ratio from comparison with the data. These lines agree well with the majority of the DMS data points (filled black circles with error bars). This suggests that using the gravitational field from the fitted exponential disk instead of the numerically computed gravitational field from the observed surface brightness is a minor concern. This also applies to using Eq.~\ref{eqn:m13b} instead of the measured data points.\\

\item For the second case, DM is still neglected, but the Jeans equation (Eq.~\ref{eqn:sig*}) is used to find the model velocity dispersions from the true double exponential stellar disk gravitational field. Thus the second term of Eq.~\ref{eqn:radder} is implicitly accounted for. These model vertical velocity dispersions are then used to find the best-fit mass-to-light ratio. The blue lines corresponding to this scenario identify mass-to-light ratios that are typically 0.1 higher than the violet lines. This, however, would assume that disks are maximal, which they cannot be because of the need to simultaneously fit the rotation curve (with an inclination chosen to be compatible with the luminous Tully-Fisher relation as in the DiskMass study).\\

\item For the third scenario, the model vertical velocity dispersions take into account the neglected terms in the Jeans equation and the DM halo that M13a used to fit the rotation curves and the stellar disk gravitational field (given a mass-to-light ratio). Those rotation curve fits assumed the mass-to-light ratios found by M13a. In our fit, the stellar mass-to-light ratio is a free parameter. Again, the model vertical velocity dispersions are computed from the true double exponential stellar disk gravitational field and the fixed DM halo. Now there are 14 out of 30 galaxies that have stellar mass-to-light ratios lower than 0.11 and only 3 galaxies have stellar mass-to-light ratios greater than 0.4.

These recovered stellar mass-to-light ratios are significantly lower than those found by the DiskMass Survey (M13a) because the dark matter gravitational field is in many cases the dominant contributor to the stellar vertical velocity dispersion. This is not unexpected since, even in the galactic mid-plane between $R=h_R$ and 3$h_R$, the dark matter density is typically only a factor of 1.5 - 3 lower than the stellar density, and declines more slowly with height. We note, however, that these new lower mass-to-light ratios are only {\it upper limits} because they do not take into account the fact that the DM halo itself now has to be made more massive to still fit the rotation curve.\\

\item The last case considered includes the rescaling of the DM halo mass necessary to still fit the rotation curve after reducing the mass-to-light ratio. For this, we consider the contribution to the total circular velocity from stars and DM at $2.2h_R$, $V_\star$ and $V_{DM}$, and transfer the loss in $V^2_\star$ (from reducing the mass-to-light ratio) to the mass of the DM halo. This factor is then used to linearly increase the vertical gravitational field from the DM halo. The process is iterated such that the final $M/L$ ratio (dashed black line in Fig.~\ref{fig:ml}) is $M/L_{i \rightarrow \infty}$, with
\beq
(M/L)_i = (M/L)_{i-1}  -  \frac{[\frac{V_\star}{V_{DM}} ((M/L)_{i-2}-(M/L)_{i-1})]^2 }{(M/L)_0}
,\eeq
where $(M/L)_0$ is the original DiskMass mass-to-light ratio (filled circles in Fig.~\ref{fig:ml}) and $(M/L)_1$ the one from case 3 above. This iteration is not done for the galaxies where $(M/L)_{1}>(M/L)_{0}$ since their dark matter contribution should actually be reduced, nor for UGC~8196 (number 30 in Fig.~\ref{fig:ml}) for which the iteration does not converge. The drop compared to case 3 is nevertheless relatively minor, as can be seen in Fig.~\ref{fig:ml} by comparing the red and dashed black lines. The median stellar mass-to-light ratios for cases 3 and 4 are 0.19 and 0.18, respectively, and the means are 0.2 and 0.19.

\end{enumerate}

The uncertainties on the best-fit stellar mass-to-light ratios are assumed to be roughly equivalent in size to those of M13a since the data, and not the method, are still the dominant source of error. To check the impact of the chosen radial interval for comparing the measured and model stellar velocity dispersions, the best-fit mass-to-light ratio in each scenario was also computed in the interval [1-3]$\times h_R$, [0.75-2]$\times h_R$ in addition to the default [0.5-3]$\times h_R$. There was little difference, suggesting the radial interval for comparison does not meaningfully affect the results.

\subsection{Shape of the dark matter halo}
In the above analysis, we have focused exclusively on spherical dark matter halos. There is, however, evidence from N-body simulations on cluster scales that halos have significant departures from spherical symmetry (e.g. \citealt{frenk88}). \cite{bullock02} found that Milky Way-scale halos were slightly less flattened than their cluster-scale equivalents. More recent studies using the Aquarius simulation (\citealt{veraciro13}) found evidence for oblate halos, but showed that the degree of non-sphericity varies with redshift and radial shell, as well as from halo to halo.

In our case, the presence of oblate halos (elongated along the disk plane) would compound the problem, but prolate (elongated in the vertical direction) halos could allow the mass-to-light ratios to increase by lowering the dark matter mass density in the disk midplane. To test the strength of this effect, we followed the strategy of  \cite{helmi04}, who stretched the {\it potential} in the vertical direction by a factor $q$. This means the pISO gravitational potential becomes

\beq
\protect\label{eqn:prolate}
\Phi(R,z)=2\pi G \rho_0 r_c^2\left[\ln\left(1+\left({r' \over r_c}\right)^2\right)+2{r_c \over r'}arctan\left({r' \over r_c}\right) \right],
\eeq
where $r'=\sqrt{R^2+{z^2\over q^2}}$.

 We do not stretch the mass distribution because this would modify the rotation curve (i.e. the gravity profile in the disk midplane). The drawback of our approach is that the density variation, as a function of radius and height above the disk, relative to the spherical halo, is not straightforward. However, in general the average density is increased in proportion to the potential scaling. \cite{helmi04} stated that stretching the potential by factors $q=1.125$ and 1.25 is roughly equivalent to average density profile axis ratios of $\tilde{q}=1.25$ and 1.67, respectively.
 
\begin{figure}
\includegraphics[angle=0,width=8.50cm]{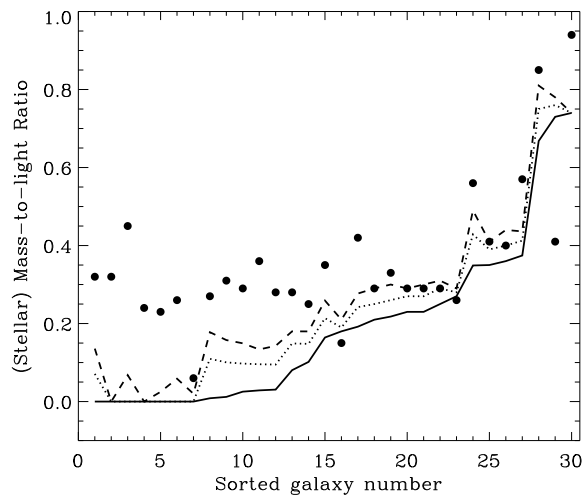}
\caption{Derived stellar mass-to-light ratios for case 4 using different shapes of dark matter halos. As per Eq~.\ref{eqn:prolate}, we use prolate halos with $q=1$, 1.125 and 1.25 (solid, dotted and dashed line, respectively). The original data points from Martinsson et al. (2013a) are given for reference.}
\label{fig:prolate}
\end{figure}

 In Fig.~\ref{fig:prolate} we compare the mass-to-light ratios derived from equivalents of case 4 for $q=1$, 1.125 and 1.25 (solid, dotted, and dashed lines, respectively). A significantly prolate halo, with $q=1.25$ can increase the mass-to-light ratios of certain galaxies usually by between 0.05 and 0.1 and at most 0.15. However, it still leaves more than 20 out of 30 galaxies with $M_\star/L_K < 0.3$ and 6 galaxies with $M_\star/L_K < 0.1$.

\subsection{Cross-terms in the Jeans equation}
Finally, we checked the influence of the neglected cross-term $\sigma^2_{Rz}$ in the Jeans equation (see e.g. \citealt{kuijken89} Eq~4). This cross-term is of course zero for a pure disk potential. \cite{kuijken89} suggested that an upper limit for the influence of the cross-term can be found by considering a spherical potential, which would be the case for the ultra-low stellar mass-to-light galaxies found here.

\begin{figure}
\includegraphics[angle=0,width=8.50cm]{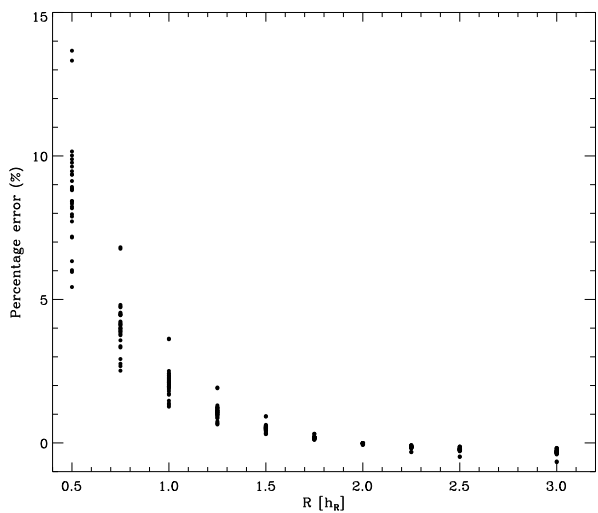}
\caption{Percentage error upper limit as a function of radius in the derived stellar vertical velocity dispersion by either neglecting the cross-terms in the Jeans equation, or including them. This is shown for all 30 galaxies assuming zero stellar mass-to-light ratios and the dark matter halo profiles fitted by Martinsson (2013a).}
\label{fig:tilt}
\end{figure}

We solved their Eq.~50 to find the stellar vertical velocity dispersions for all 30 galaxies, with the stellar-mass-light ratio set to zero. In Fig.~\ref{fig:tilt} we compare this stellar vertical velocity dispersion with the one found ignoring the cross-term. The percentage errors suggest that for $R = 0.5h_R$ including the cross-terms decreases the model velocity dispersions by between 5 and 14\%, thus this would slightly increase the fitted stellar mass-to-light ratios. For  $R \ge 0.75h_R$ the percentage errors are lower than 8\% and for $R > h_R$ they are basically irrelevant relative to the much larger observational errors.

\section{Conclusions and discussion}
\protect\label{sec:conc}
Vertical stellar velocity dispersions can be a powerful tool to estimate the stellar mass-to-light ratio in the disks of nearly face-on galaxies, and they do not suffer from the disk-halo degeneracy. However, the observations of such external galaxies are, in principle, now accurate enough to require modelling beyond the zeroth-order approximation, which consists of using Eq.~\ref{eqn:vdk}. Here we recalled the well-known problem of using Eq.~\ref{eqn:vdk} to estimate the surface density of a galaxy, both because it neglects the radial derivative term of the Poisson equation and because the DM gravitational field can provide an important contribution to the stellar velocity dispersions, especially when the disks are sub-maximal, as the DiskMass survey results imply.

The main point of this article was to show that the small, measured vertical velocity dispersions of the DiskMass survey lead to an iterative reduction of the mass-to-light ratio. Assuming no DM, they imply rather low stellar mass-to-light ratios, but this in turn requires heavy DM halos to fit the rotation curves. The vertical gravitational field of the DM then contributes to the vertical stellar velocity dispersion, meaning that the stellar mass-to-light ratio must be decreased again. This moreover requires an even heavier DM halo to fit the rotation curve. Once the process converges, the actual stellar mass-to-light ratios drop to a median value of 0.18, or 0.19 on average. This is significantly lower than the 0.3 value originally found by the DiskMass Survey in the absence of DM and the 0.24 found when the DM contribution was subtracted in a manner that was not fully self-consistent. 

Prolate dark matter halos, possible from cosmological simulations, would decrease the dark matter density in the mid-plane. We stretched the potential in the vertical direction by a factor $q=1.125$ and 1.25 and found that using $q=1.25$, which is roughly equivalent to an average density profile axis ratio of $\tilde{q}=1.67$, made only a small difference to the mass-to-light ratios - a typical increase of between 0.05 and 0.1. This still leaves more than 20 out of 30 galaxies with mass-to-light ratios lower than 0.3 and 6 with ratios lower than 0.1. Extreme prolate halos ($q>1.5$) could play a more significant role in increasing the mass-to-light ratios.

We checked that the cross-terms in the Jeans equation were insignificant at radii beyond $h_R/2$ using the method of \cite{kuijken89}.

We note that the K-band stellar mass-to-light ratios appear to be significantly non-uniform from one galaxy to another, in stark contrast with predictions from population synthesis models. For instance, 14 galaxies have stellar mass-to-light ratios lower than 0.11, 12 of which are lower than 0.03. The bulk of the rest are between 0.16 and 0.38, but the highest is 0.74. In addition to the variation, the absolute values of the stellar mass-to-light ratios are also surprisingly low. Indeed, adopting the V band as a reference point - based on Milky Way stellar counts - \cite{mcgaugh14} showed that to yield self-consistent galactic stellar masses in different bands, the value $M_{\star}/L_K = 0.6$ was preferred. This value is generally consistent with maximal disks, which is also consistent with the Milky Way dynamical analyses \citep{bovy13}. It would nevertheless require a $>$30\% increase of the stellar velocity dispersions measured by the DiskMass survey.

The present results could suggest that galaxies are remarkably non-uniform in terms of their stellar mass-to-light ratios and that they are in stark conflict with the models of stellar population synthesis, or that the DM halos are highly non-spherical and
therefore do not contribute much to the vertical gravitational field close to the disk while allowing for a high circular velocity. Alternatively, the measured stellar velocity dispersions of the DiskMass survey might have been under-estimated by $>$30\%.

\section{Acknowledgements} GWA is a postdoctoral fellow of the FWO Vlaanderen (Belgium). We thank the referee for several important suggestions that improved the manuscript.

\end{document}